\documentclass[12pt]{article}
\pdfoutput=1

\usepackage{jheppub} 

\usepackage{amssymb,amsbsy,amsfonts,amsmath}
\usepackage{epsfig}
\usepackage{cancel}
\usepackage{color}

\usepackage{subfigure}
\usepackage{graphicx}
\usepackage{epstopdf}
\usepackage{float}

\def\bea{\begin{eqnarray}}
\def\eea{\end{eqnarray}}
\def\be{\begin{equation}}
\def\ee{\end{equation}}
\def\beqn{\begin{eqnarray}}
\def\eeqn{\end{eqnarray}}
\def\beq{\begin{equation}}
\def\eeq{\end{equation}}

\def\Dslash{\not{\hbox{\kern-4pt $D$}}}
\def\pslash{\not{\hbox{\kern-4pt $p$}}}

\newcommand{\ie}{\textit{i.e.}}
\begin{document}

\baselineskip=21pt \pagestyle{plain} \setcounter{page}{1}

\vspace*{-1.7cm}

\begin{flushright}{\small FERMILAB-PUB-15-535-T}\end{flushright}

\vspace*{0.2cm}
 
 \begin{center}

{\Large \bf  Dark matter beams at LBNF
} \\ [9mm]

{\normalsize \bf Pilar Coloma, Bogdan~A.~Dobrescu,  Claudia~Frugiuele, Roni Harnik } \\ [4mm]
 {\it Theoretical Physics Department, Fermilab, Batavia, IL 60510, USA } 

\vspace*{0.5cm}

December 11, 2015

\vspace*{0.7cm}

{\bf \small Abstract}

\vspace*{0.4cm}

\parbox{14.8cm}{

High-intensity neutrino beam facilities may  produce a beam of light dark matter when protons strike the target. 
Searches for such a dark matter beam using its scattering in a nearby detector must overcome the large neutrino background. We characterize the spatial and energy distributions of the dark matter and neutrino beams, focusing on their differences to enhance the sensitivity to dark matter. We find that a dark matter beam produced by a $Z'$ boson in the GeV mass range is both broader and more energetic than the neutrino beam. The reach for dark matter is maximized for a detector sensitive to hard neutral-current scatterings, placed at a sizable angle off the neutrino beam axis. 
In the case of the Long-Baseline Neutrino Facility (LBNF), a detector placed at roughly 6 degrees off axis and at a distance of about 200 m from the target would be sensitive to $Z'$ couplings as low as 0.05. This search can proceed symbiotically with neutrino measurements.
We also show that the MiniBooNE and MicroBooNE detectors, which are on Fermilab's Booster beamline, happen to be at an optimal angle from the NuMI beam  and could perform
searches with existing data. 
This illustrates potential synergies between LBNF and the short-baseline neutrino program if the detectors are positioned appropriately.
}

\bigskip\bigskip

\end {center}

\newpage

\tableofcontents

\section{Introduction}

Dark matter (DM) provides solid evidence for physics beyond the Standard Model (SM), but its particle nature remains unknown.
A central question is whether DM particles experience interactions with ordinary matter beyond gravity. 
Direct detection experiments \cite{Cushman:2013zza} 
have imposed impressive constraints on the interactions between nucleons and DM particles of mass larger than about 5 GeV. 
These experiments lose sensitivity quickly at lower masses because light dark matter particles moving at the viral velocities of our galactic halo 
would yield very low recoil energies in collision with nuclei or atoms. 
Interactions of DM with quarks or gluons are also explored at high-energy colliders, for example through monojet searches \cite{Goodman:2010yf,
Goodman:2010ku,
Rajaraman:2011wf,
Bai:2010hh,
Fox:2011pm,
Aaltonen:2012jb,
Khachatryan:2014rra,
Aad:2015zva}. 
If these
interactions are due to  a light mediator, however, the collider searches are less sensitive.

Therefore, the question of how to conduct light dark matter searches is urgent and compelling.
A potentially promising direction is to use proton fixed-target experiments to probe DM couplings to quarks~\cite{Batell:2009di, deNiverville:2011it, deNiverville:2012ij, Dharmapalan:2012xp, Batell:2014yra, Dobrescu:2014ita} (other proposals for light DM searches  have been explored in
\cite{Reece:2009un, 
Bjorken:2009mm, 
Essig:2010xa, 
Essig:2010gu, 
Izaguirre:2013uxa, 
Morrissey:2014yma, 
Izaguirre:2014dua, 
Soper:2014ska,
Kahn:2014sra, 
Mohanty:2015koa,
Gardner:2015wea,
Alekhin:2015byh}).
An interesting type of mediator is a leptophobic $Z'$ boson. For a $Z'$ mass in the $\sim$1--10  GeV range, 
the limits on its coupling to quarks are remarkably loose \cite{Dobrescu:2014fca}.  
A dark matter beam originating from the decay of a leptophobic $Z'$, produced by protons accelerated in the Booster at Fermilab, may lead to a signal 
in the MiniBooNE experiment  \cite{Batell:2014yra}. This signal decreases fast for $M_{Z'}$ above 1 GeV, because the Booster proton energy is only 8 GeV. 
By contrast, protons accelerated at 120 GeV in the Main Injector scattering off nucleons may produce a leptophobic $Z'$ as heavy as $\sim 7$ GeV, and the DM particles originating in the $Z'$ decay may lead to neutral-current events in neutrino detectors \cite{Dobrescu:2014ita}.

Here we analyze the sensitivity of neutrino detectors to the DM beam produced in leptophobic  $Z'$ decays. We focus on a high-intensity
proton beam of $\sim \! 100$ GeV,  as that proposed at the Long-Baseline Neutrino Facility~\cite{lbnf} (LBNF). 
We consider deep-inelastic neutral-current scattering as the main signal. The challenge of using neutrino facilities to look for a DM beam is that neutrino events represent an irreducible background. In~\cite{Batell:2014yra} it is  proposed to conduct a special run of the beam in which the magnetic horns are turned off, leading to a more dilute neutrino beam. Here we will take a different approach, namely to exploit the difference between the dark matter and a focused neutrino beam and consider a detector that is located accordingly. This search for dark matter does not disrupt the normal neutrino research program.

More specifically, we will see that the signal and main background contributions have very different energy and angular profiles, which can be exploited to enhance the signal significance. We perform a simple optimization study using the signal significance in order to determine the optimal position of a detector.  
We determine that an angle of approximately 6 degrees with respect to the decay pipe direction would maximize the sensitivity. 
Applying these results to the NuMI beamline, we find that 
the NOvA near detector, in spite of being located slightly off-axis, does not provide a sufficient suppression of the neutrino background. 

The paper is structured as follows. Section~\ref{sec:model} reviews the main features of the model considered. In Section~\ref{sec:sig-bg} we discuss the main differences between the DM signal and the neutrino background, paying special attention to their energy and angular distributions. In Section~\ref{sec:location} we identify the optimal off-axis location for a detector, based on the signal-to-background expected ratio, and the $\chi^2$ sensitivity contours for two close-to-optimal locations are presented. Our conclusions are 
presented in Section~\ref{sec:concl}. The computation of the neutrino flux due to kaon decays is outlined in the Appendix.

\section{Dark matter and a light $Z'$ boson}
\label{sec:model}\setcounter{equation}{0}

We consider a $Z'$ boson, associated with the $U(1)_B$ gauge group, which couples to the quarks $q = u,d,s,c,b,t$ and to a dark matter fermion $\chi$:
\be
\mathcal{L}_q = \frac{g_z}{2}  Z^{\prime}_\mu \left(  \frac{1}{3} \sum_q  \; \overline q \gamma^\mu q  + z_\chi \overline \chi \gamma^\mu \chi \right)
  ~~,
\label{eq:Lagrangian}
\ee
where  $g_z$ is the gauge coupling, and  $z_\chi$ is the  $U(1)_B$ charge of $\chi$.
In the case where $\chi$ is a complex scalar, the $\overline \chi \gamma^\mu \chi$ term in Eq.~(\ref{eq:Lagrangian})
is replaced by $i \chi \partial^\mu \chi + {\rm H.c.}$

The ratio of decay widths into $\chi's$ and quarks is 
\be
\frac{\Gamma( Z' \rightarrow  \chi \bar \chi ) }{\Gamma( Z' \rightarrow  q\bar q )}
=   \,  \frac{3 z_{\chi}^2}{N_f(M_{Z'})}  \;  F_\chi (m_{\chi}/M_{Z'}) ~~ ,
\ee
where $\Gamma( Z' \rightarrow  q\bar q )$ stands for the sum over the partial decay widths into all quarks, $N_f (M_{Z'})$ is the effective number of quark flavors of mass below $M_{Z'}/2$,
and the function $F_\chi $ is defined by 
\be
F_\chi  (x) =  \left\{ 
\begin{array}{c}   \Big(1+2 x^2 \Big)  \Big(1-4  x^2  \Big)^{\! 1/2}   \; ,  \;\; \;   {\rm if \; \chi \; is \; a \;  Dirac \;  fermion \; }   ~,
\\ [3mm]  {\displaystyle \frac{1}{4}} \Big(1-4  x^2  \Big)^{\! 3/2}   \; ,  \;\; \;   {\rm if \;  \chi \; is \;  a \: complex \;  scalar \; }   ~.
 \end{array} \right.
 \ee
The effective number of quark flavors below $M_{Z'}/2$ takes into account the phase-space suppression for $Z'$ decays into hadrons, and thus is not an integer.
For $M_{Z'}$ in the $\sim$ 1--3.7 GeV range, $N_f (M_{Z'}) \approx 3$, while for $M_{Z'}$ in the $\sim$ 3.7--10 GeV range, $N_f (M_{Z'}) \approx 4$,
with large uncertainties for $M_{Z'}$ near the $s\bar s$ and $c\bar c$ thresholds.

The existing constraints on the $Z'$ coupling in the 1--10 GeV mass range are rather weak, given that this is a leptophobic boson:
\begin{enumerate}
\item $Z'$ exchange induces invisible decays of quarkonia with a branching fraction 
\cite{Graesser:2011vj}:
\bea
B(J/\psi \rightarrow \chi \bar \chi ) =  \frac{ 4 g^4_z z_\chi^2}{ g^4 \sin^4\!\theta_W} \,   \left( 1 - \frac{M^2_{Z'}}{M^2_{J/\psi} }\right)^{\! -2} 
F_\chi(m_{\chi}/M_{J/\psi})  \, B( J/\psi \rightarrow \mu^+ \mu^-) \;\; ,
\eea
if $m_\chi < M_{J/\psi}/2$, and the analogous expression for $\Upsilon$.
The  $ 90 \% $ confidence level (C.L.) limits on invisible branching fractions are 
$B ( J/\psi \rightarrow  \chi \bar \chi) < 7 \times 10^{-4}$ \cite{Ablikim:2007ek},
and
$B ( \Upsilon \rightarrow  \chi \bar \chi) < 3 \times 10^{-4}$ \cite{Aubert:2009ae}. 

\item A kinetic mixing between the $Z'$ boson and the photon, $  - (\epsilon_B/2) Z'_{\mu \nu} F^{\mu \nu}$, arises at one loop 
with \cite{Graesser:2011vj}:
\be
\epsilon_{B} \sim 10^{-2} g_z \;  
\ee
at the 10 GeV scale. As a result, the $Z'$ boson can be produced in $e^+e^-$ collisions, albeit with a very small rate. 
The BaBar limit \cite{Aubert:2008as} on  $\Upsilon(3S)$ decay into a photon and missing energy has been reinterpreted \cite{Essig:2013vha} as 
a limit on $e^+e^- \to \gamma Z'$ with the $Z'$ produced through its kinetic mixing. This limit is competitive with the one from 
$\Upsilon \to \chi \bar \chi $ decay only for $M_{Z'}$ in the 4.6--5 GeV range.

\item Monojet searches \cite{Aaltonen:2012jb} at hadron colliders set a bound on $g_z $   \cite{Shoemaker:2011vi}
 \be
  g_z^2 \, B( Z' \rightarrow \chi \bar \chi)  <  
  \, 1.4 \times 10^{-2} ~~.
\ee
This limit is almost independent of $M_{Z'}$ in the range considered here, and it is weaker than the limit from $\Upsilon \to \chi \bar \chi $. 

\item There is also a limit on $g_z$ from the requirement that the $U(1)_B$ gauge symmetry is anomaly free, which follows from the collider limits on 
anomalons (the new electrically-charged fermions that must cancel the gauge anomalies) \cite{Dobrescu:2014fca}. 
This limit is rather stringent for $M_{Z'} \lesssim 3$ GeV for a minimal set of anomalons, but with a larger anomalon set it becomes looser 
than the one from invisible $J/\psi $ decays. Here we consider the latter case.
\end{enumerate}
Overall, values of the gauge coupling $g_z$ as large as of order 0.1 are allowed for $M_{Z'}$ in the 1 -- 10 GeV, with the exception 
of small regions near the  $J/\psi$ and $\Upsilon$ masses.
We will plot the strongest limits on $g_z$ as a function of $M_{Z'}$ in Section 5 (Fig.~\ref{fig:chi2}). 
We will not discuss possible cosmological constraints on the parameter space  which  arise when $ \chi $ is the dominant form of dark matter.
Possible viable dark matter scenarios are discussed in  \cite{Dobrescu:2014ita}.

\bigskip

\section{Neutrinos versus dark matter  at fixed target experiments}
\label{sec:sig-bg}

The search for a dark matter beam in a neutrino facility must deal with the neutrino background.  To mitigate this, new physics searches need to be tailored to maximize the signal to background ratio (or the signal significance), by looking for particular signals and in particular regions of phase space. It is convenient to separate the production, which occurs mostly in the target, from detection, which takes place in a distant detector. In this section we discuss the production and detection mechanisms both for neutrinos and dark matter, emphasizing the main differences between them. 

\subsection{ Production mechanisms for dark matter and neutrinos }
\label{subsec:production}

In the model considered in this work, the dark matter is pair produced via the decay of a $Z'$ boson, of mass in the GeV range, resonantly produced in the target by proton
scattering off nucleons, 
\be
q\bar q \to Z' \to \chi \bar \chi\,.
\ee
Searching for  mediators in the GeV range requires the use of energetic beams (in NuMI the protons have 120 GeV), and the production cross section is much smaller than
 that for mesons.\footnote{This is different from the situation considered in \cite{Batell:2014yra} where the focus is on a $Z'$ with $M_{Z'}\sim \mathcal{O}(100)$~MeV. There the $Z'$ is  produced via meson decays, at a much higher rate. The lighter $Z'$ mass and the efficient production mechanism allows to search for such mediators using low-energy proton collisions, such as those occurring in the Booster beamline target (where the incident proton energy is 8 GeV). }
This means that the problem of reducing the neutrino backgrounds produced in meson decays is nontrivial. 

Let us consider a $Z'$  of mass $M_{Z'}$ that is produced in the target with an energy $E_{Z'}$.
The energy of the dark matter particle $\chi$ in the final state can be derived from 2-body kinematics. In the lab frame it reads:
\bea
E_{\chi}=  
\frac{M^2_{Z'}}{2 E_{Z'} (1-\beta \cos{\theta})}
\label{eq:Echi}
\eea
where $\beta$ is the $Z'$ velocity, $\theta$ is the angle between the $\chi$ and $Z'$ momenta, and we have neglected the mass of the dark matter assuming that it is much smaller than the $Z'$ mass. Since the transverse momentum of the initial $q\bar q$
system is small (we are only considering production at leading order), the $Z'$ is produced in the forward direction. 
As a result the angle of the dark matter with respect to the decay pipe can be directly identified with $\theta$. 

As will be discussed in detail in Sec.~\ref{subsec:detection}, the main background is due to  very energetic neutrinos reaching the detector. For neutrinos produced in meson decays, a similar relation as in Eq.~(\ref{eq:Echi}) holds between the meson and neutrino energy, just replacing the $Z'$ variables with the parent meson variables, and $E_\chi \rightarrow E_\nu$. Thus, in the case of pions, neutrinos emitted with a sizable angle have very low energies regardless of the parent pion energy because of the low pion mass in the denominator. 
This fact, which is exploited both in the T2K and NOvA experiments to get a narrow neutrino spectrum at low neutrino energies, will also be beneficial in our case to reduce the neutrino background at high energies. For off-axis angles larger than 2 degrees no significant number of energetic neutrinos coming from pion decays would reach the detector, assuming a (relatively well) collimated pion beam. We may henceforth consider only angles above 2 degrees and ignore backgrounds from pion decay.

\begin{figure}[t]
 \begin{center}
  \includegraphics[width=.65\textwidth]{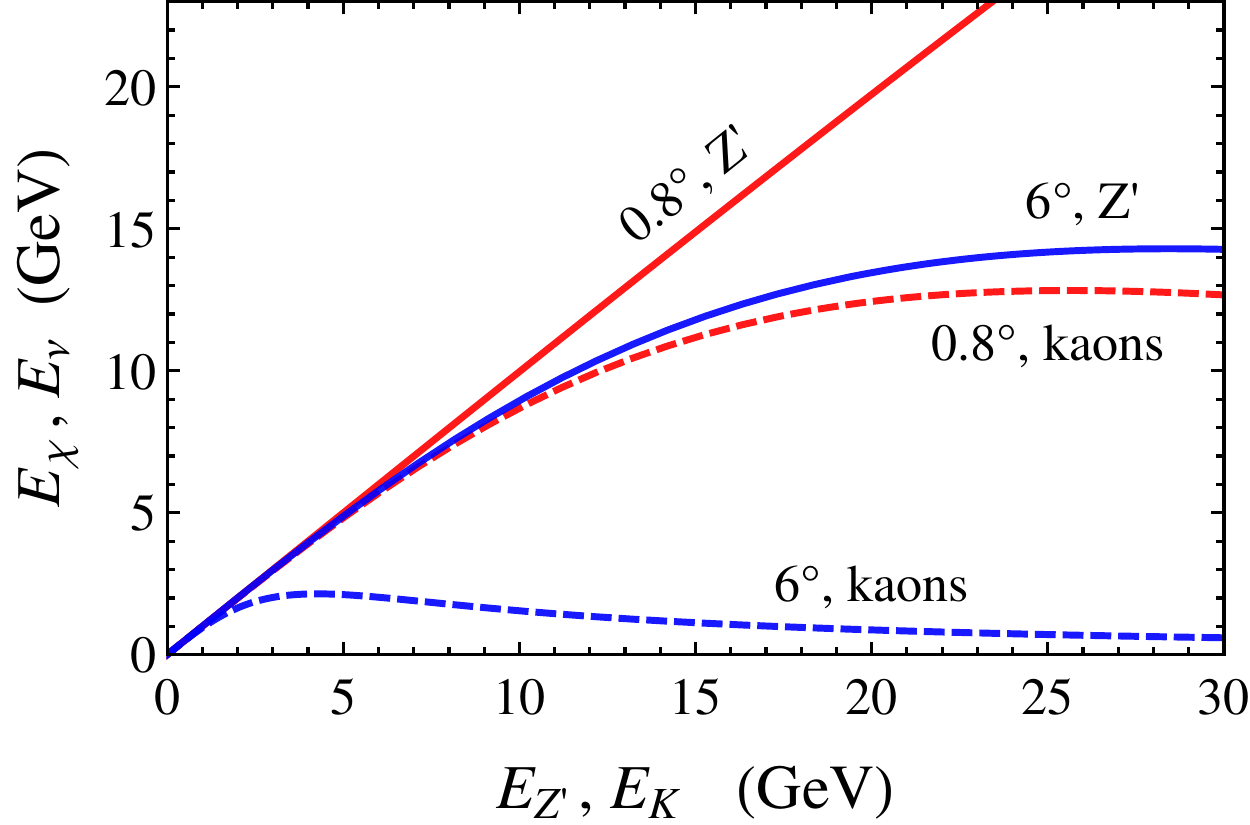} 
     \end{center}
     \caption{ Energy of the produced particle as a function of the decaying particle energy.
       The case of DM particles produced in $Z'$ decays is shown as solid lines, and the case of neutrinos produced in kaon decays  
is shown as dashed lines.
The results are shown for two different off-axis angles: $0.8^\circ$ (red) and  $6^\circ$ (blue), matching approximately the off-axis angles 
(seen from the target) of the NOvA near detector and the MiniBooNE detector with respect to the NuMI beamline.}
\label{fig:energydist}
\end{figure}

Following the above argument, it is clear that our main background is going to come from neutrinos produced in kaon decays, which will lead to a more energetic flux of neutrinos off axis. Nevertheless, since $M_K \ll M_{Z'}$, the resulting neutrino flux will still be much less energetic than the dark matter flux. This can be understood from Eq.~(\ref{eq:Echi}) and is illustrated in Fig.~\ref{fig:energydist}, where the energy of the daughter particle is shown as a function of the parent energy, both for $Z'$ and kaon decays. The results are shown for two different off-axis angles, which roughly correspond to the angles subtended by both the NOvA near detector and the MiniBooNE detector, measured with respect to the NuMI beamline.

So far we considered the decay of a $Z'$ boson or a kaon produced with a given energy. This qualitative understanding must be folded with their respective energy distributions as they exit the target. In order to compute the dark matter energy profile, we generate proton-proton collisions using \texttt{MadGraph/MadEvent 5}~\cite{Alwall:2011uj} with \texttt{NNPDF23LO1} parton distribution functions (PDFs)~\cite{Deans:2013mha}. The implementation of the model into \texttt{MadGraph} has been done using the \texttt{FeynRules} package~\cite{Christensen:2008py}. The LHE files have been parsed using \texttt{PyLHEF}~\cite{pylhef}. Due to the short baselines considered for this setup, in the 100--700 m range, the size of the detector will also have an impact on the energy profile. For simplicity, we consider a generic spherical detector of a similar size to the MiniBooNE detector~\cite{AguilarArevalo:2008qa} (a radius $R_{det}=6.1$~m, and a mass of 800~tons). 

\begin{figure}[t]
 \begin{center}
  \includegraphics[width=0.49\textwidth]{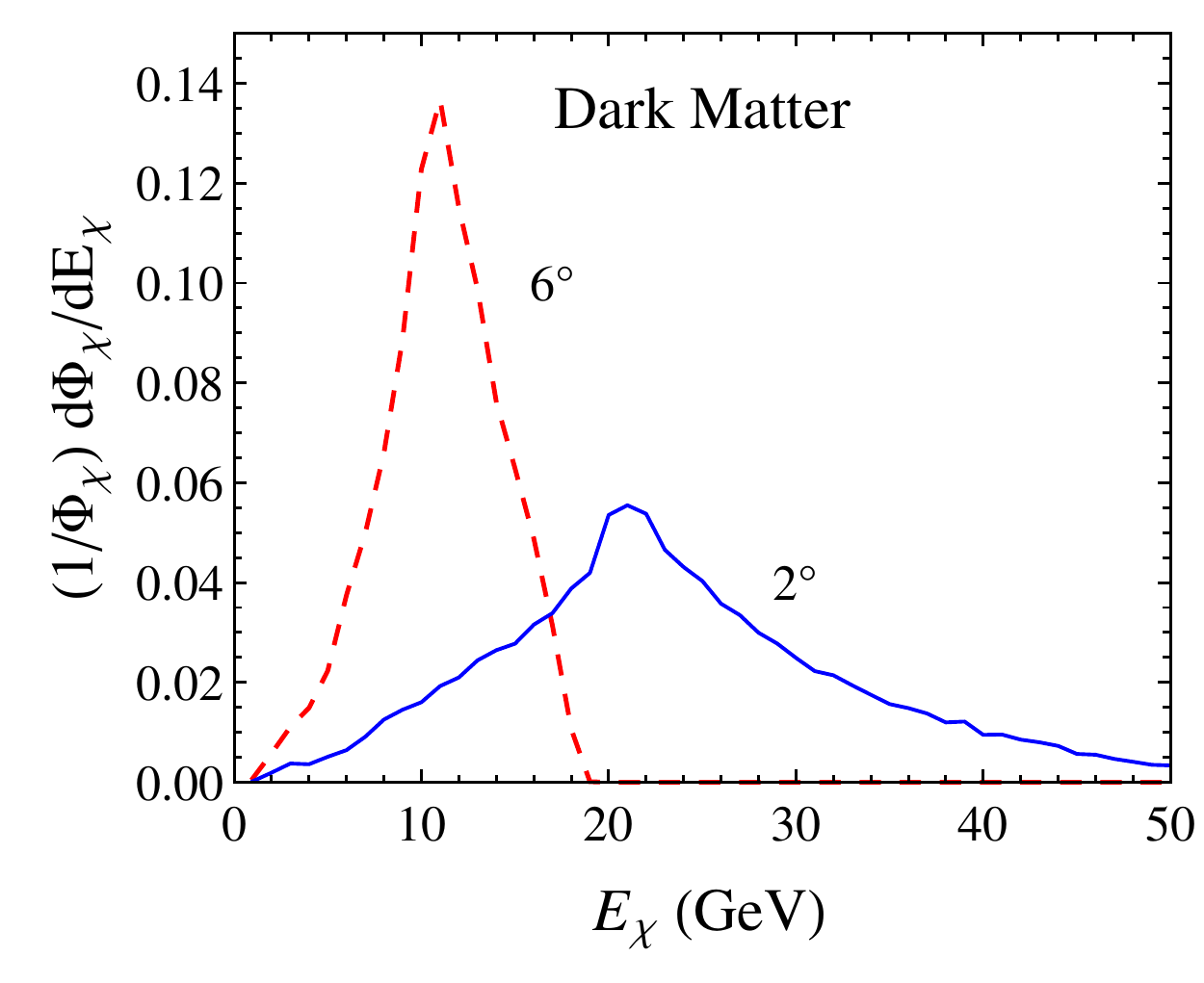} \
  \includegraphics[width=0.49\textwidth]{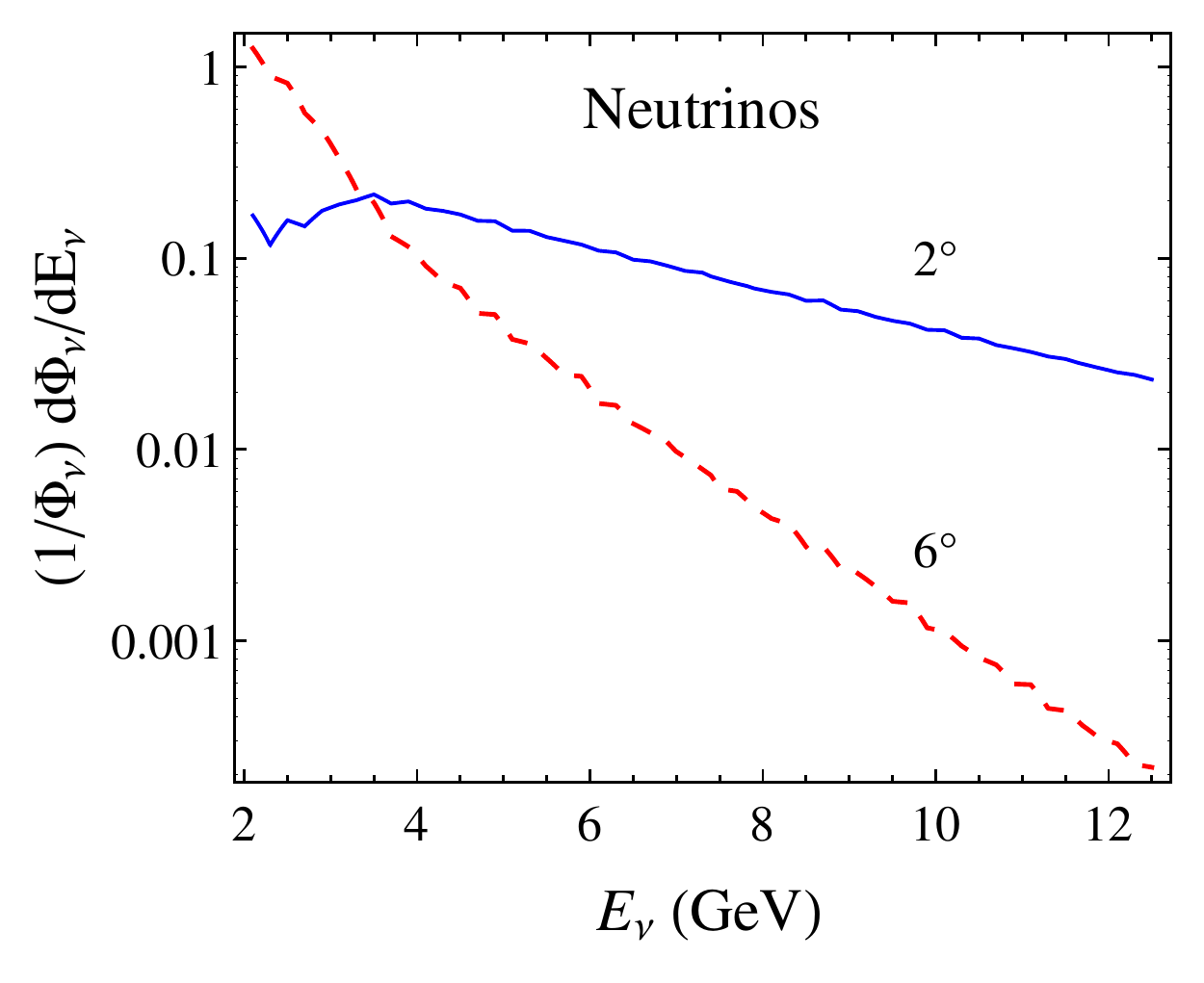} 
 \end{center}
     \caption{Differential flux that reaches a MiniBooNE-size detector located 745~m away from the target, for DM particles (left) and for neutrinos (right), 
produced from 120~GeV protons scattering off nucleons at rest. 
Results are shown for two different off-axis angles, $2^\circ$ (solid) and  $6^\circ$ (dashed).  }
\label{fig:Edm-dist}
\end{figure}

The final dark matter flux expected at the detector can be seen in the left panel of Fig.~\ref{fig:Edm-dist} for a mediator with $M_{Z'} = 3$~GeV and a fermionic dark matter candidate with $m_\chi = 750$~MeV. Results are shown for two different values of the off-axis angle $\theta$, as a function of the dark matter energy (see also Fig~5 in Ref.~\cite{Dobrescu:2014ita}). For comparison, in the right panel we show the neutrino flux as a function of the neutrino energy, for the same off-axis angles. Indeed, comparing the two panels of Fig.~\ref{fig:Edm-dist} we see that the difference in mass between a few GeV $ Z'$ and kaons (and pions) offers an interesting handle to distinguish between dark matter and neutrinos, since the latter tend to be less energetic (especially when the detector is placed off-axis). This will also provide an extra relative suppression for the background with respect to the signal, since the interaction cross section at the detector grows with the energy of the incoming particle.

We have shown that the energy spectrum of dark matter that reaches an off-axis detector is harder than the neutrino spectrum reaching it. The second important difference between production of dark matter with a GeV mediator and neutrinos from kaon decay is going to be the angular dependence of the flux. While dark matter is produced from the decay of a spin 1 particle, neutrinos are produced from a spin zero meson, which will affect the angular distribution of the particles produced in the decay. Moreover, the probability for the daughter particle to be emitted in the direction of the off-axis detector will depend on its energy. For neutrinos this probability reads
\bea
\frac{dP_{\nu}}{d\Omega} = \frac{1}{4 \pi \gamma^2(1-\beta \cos{\theta})^2 }
\label{eq:angulardist-nu}
\eea
where $\Omega$ is the solid angle in the kaon rest frame, and $\beta$ refers to the parent velocity. 
The dark matter distributions, on the other hand, will be different depending on whether $\chi$ is a fermion or a scalar particle as follows:
\bea
\frac{dP_{F,S}}{d\Omega} = \frac{ 1 \pm ( 1-  M^2 \sin^2{\theta} )  }{\gamma^2(1-\beta \cos{\theta})^2 } 
\label{eq:angulardist-dm}
\eea
where $F,S$ stand for fermion and scalar dark matter, respectively, and $M = 1/( \gamma(1-\beta \cos{\theta}))$. 

\begin{figure}[t]
 \begin{center}
  \includegraphics[width=.7\textwidth]{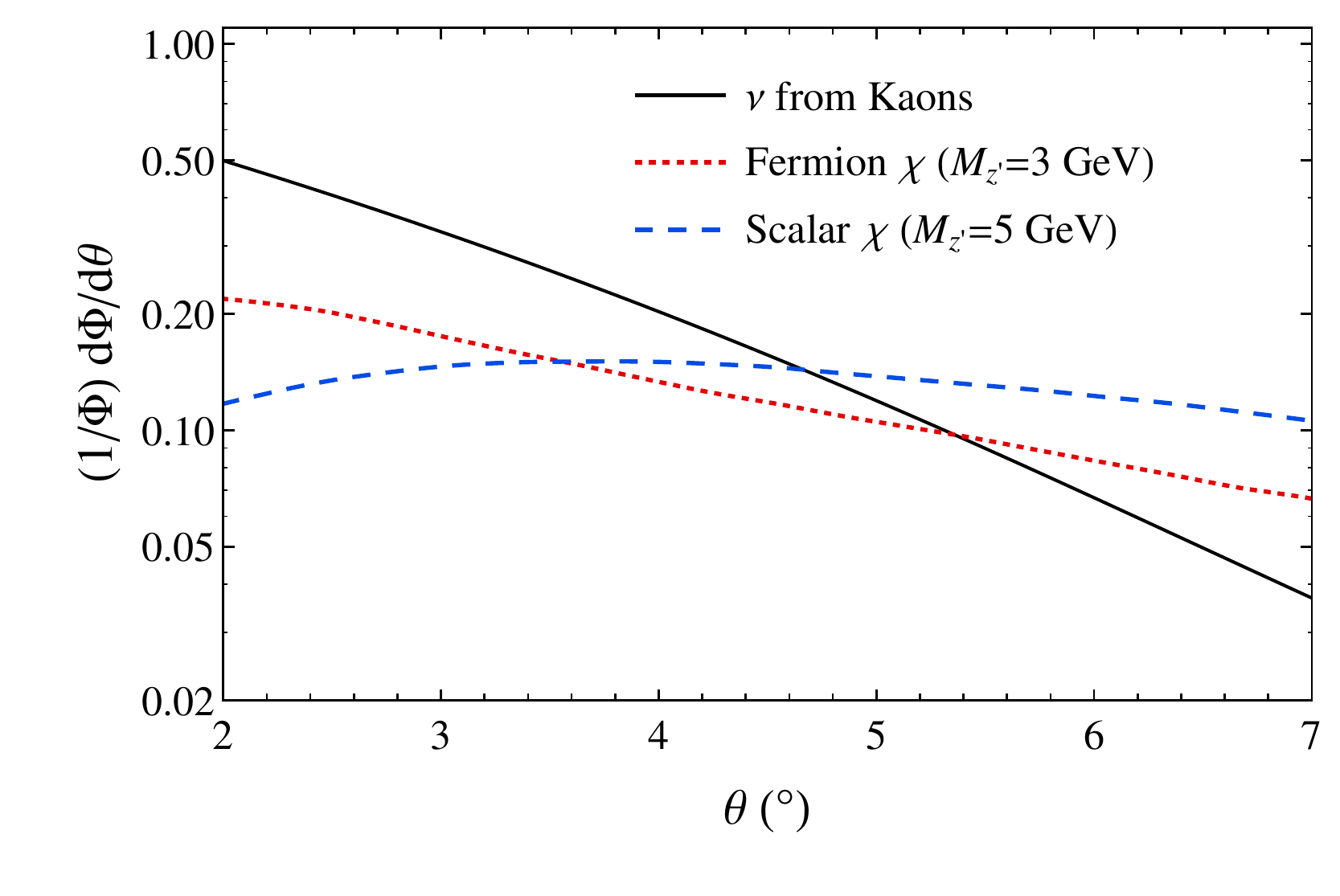} 
     \caption{    Differential flux as a function of the off-axis angle, for fermion (dotted line) and scalar (dashed line) dark matter, with $M_{Z'}$ indicated in the legend. The solid 
     line shows the neutrino flux coming from kaon decays. The angular acceptance of the detector corresponds to a MiniBooNE-size detector located at $L=745$~m from the target. }     
   \label{fig:angulardist}
 \end{center}
\end{figure}

The solid angle discussed above is defined with respect to the rest frame of the parent particle, which does not necessarily coincide with the beam axis. As already mentioned, in the case of the $Z'$ this is a negligible effect -- the $Z'$ is emitted very forward and to a good approximation its direction is the beam axis. Therefore, it is straightforward to obtain the dark matter flux as a function of its energy, by folding the $Z'$ energy distribution with Eqs.~(\ref{eq:Echi}) and~(\ref{eq:angulardist-dm}). The case of kaon decay is more complicated, though, as the kaon is typically produced with other hadrons which can balance its $p_T$. The kaon momentum thus generally subtends a non-zero angle with respect to the lab frame, which has to be accounted for when computing the neutrino flux reaching the detector. In this work, the neutrino flux has been computed using publicly available data for the kaon momenta and energy from Monte Carlo simulations of the NuMI target when exposed to 120~GeV protons~\cite{Adamson:2008qj,Pavlovic:2008zz,zarko}. More details on the computation of the neutrino flux can be found in the Appendix.  
The angular distributions with respect to the off-axis angle are shown in Fig.~\ref{fig:angulardist}, both for neutrinos coming from kaon decays and for dark matter resonantly produced via $Z'$. Since we are only interested in events producing very energetic hadron showers in the detector, these distributions have been obtained considering only particles with energies above 2 GeV. The different lines correspond to total number of neutrinos, scalar $\chi$ or  fermion $\chi$, which reach a MiniBooNE-like detector placed at $L=745$~m from the target. In all cases, the angular acceptance of the detector has been taken into account. From this figure it is evident that the suppression with the off-axis angle is stronger for the neutrino flux than for the dark matter fluxes.

\bigskip

\subsection{ Detection via neutral-current events }
\label{subsec:detection}
In the previous section we have shown that the dark matter flux tends to be more energetic than the neutrino flux at off-axis locations, and that the angular dependence of the spectrum is also different for the signal and background.   
We now evaluate if the signatures for the signal and background events in the detector are sufficiently different to allow a dark matter search at neutrino detectors.

In the model considered in this work, the dark matter particles produced at the target would give an excess of neutral-current events at the detector, which in principle may be confused with neutrino neutral-current events. Since the dark matter flux is expected to be more energetic, we consider only deep-inelastic scattering events, and we require that the energy deposited by the hadronic shower at the detector is above 3~GeV. This requirement further suppresses the neutrino contribution with respect to the dark matter signal. 

The total cross section as a function of the energy of the incident particle, as well as the hadronic energy distributions, are computed with \texttt{MadGraph} both for the neutrino and dark matter events since, in this range, the cross section can be computed within the parton model\footnote{ In order to guarantee that the parton model can be used in our case, we fix the factorization scale of the process to be equal to the mass of the $Z'$, and a conservative cut is imposed on the momentum transfer of the process, $Q^2 \gtrsim 2~\textrm{GeV}^2$. }. We have checked that the neutrino neutral-current cross section obtained with \texttt{MadGraph} is approximately $\sigma^{NC}_\nu \sim 10^{-2}~\textrm{pb}$ for neutrino energies around 10~GeV, which is in reasonable agreement with the literature, see e.g. Ref.~\cite{Formaggio:2013kya}. In the case of the dark matter, due to the much lighter mediator mass, the cross section is much larger. For instance, $M_{Z'}=3$~GeV, $g_z =0.1$ and $z_\chi = 3$ gives a dark matter NC cross section of $\sigma^{NC}_{\chi} \sim  5$~pb for $E_\chi \sim 10$~GeV. The much larger interaction cross section will provide an extra enhancement of the signal with respect to the neutrino background. 

\begin{figure}[h]
 \begin{center}
	  \includegraphics[width=.65\textwidth]{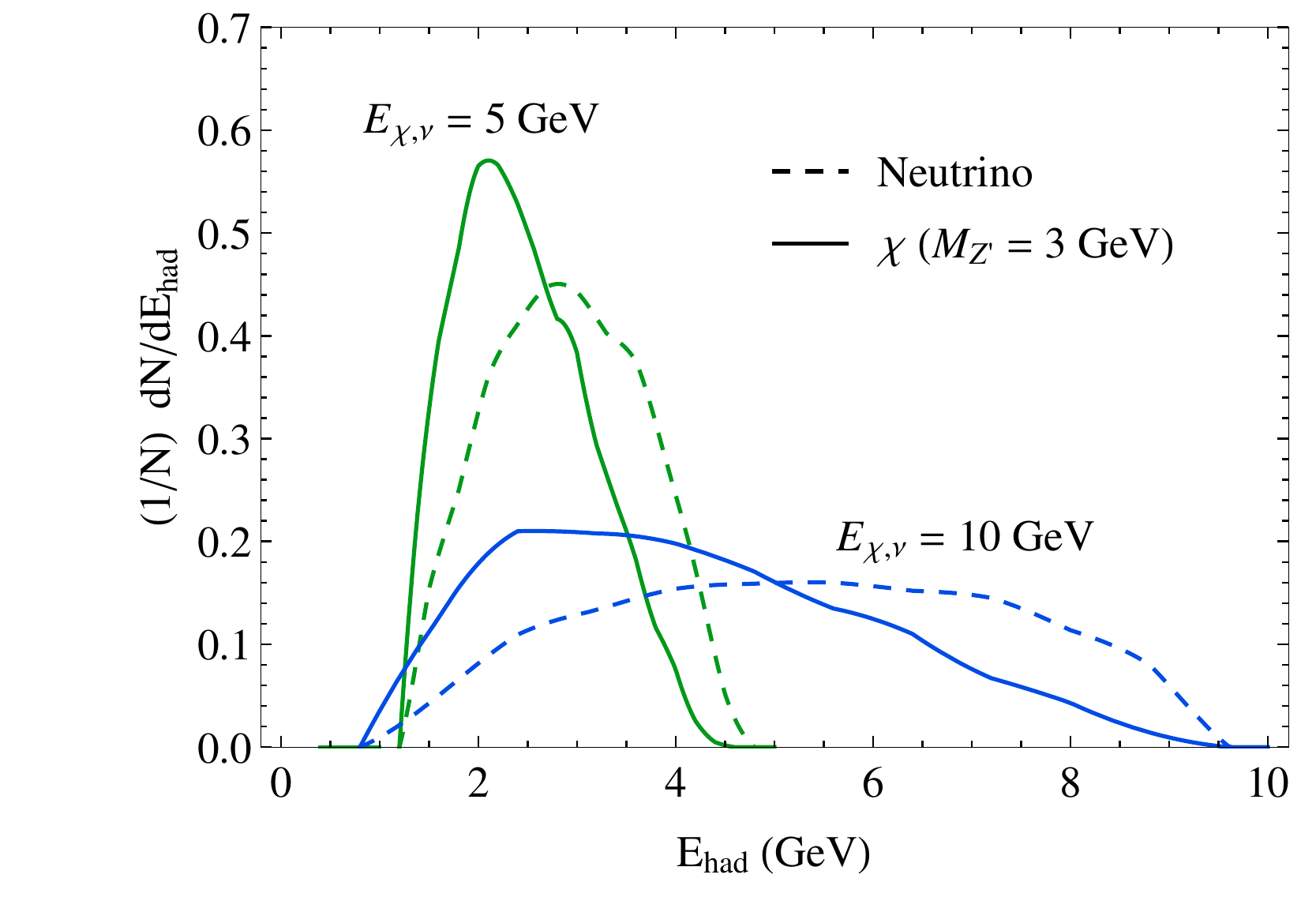}  
	\caption{ Probability to get an outgoing hadronic shower with a certain energy, for neutrino (dashed lines) and for dark matter (solid lines) neutral-current events. Two sets of lines are shown for incident particle energies of 5 GeV and 10 GeV. In the dark matter case, $M_{Z'} = 3$ GeV and $m_\chi = 750$~MeV.  }
	\label{fig:jetPDF}
 \end{center}
\end{figure}

At first sight, the kinematics of signal and background scattering  should be rather different. At the matrix element level there is a notable difference due to the small $M_{Z'}/M_Z$ ratio. The $Z'$ propagator is proportional to $(q^2-M_{Z'})^{-1}$, $q^2$ being the squared-momentum transfer. For the background, instead, $M_{Z'}$ is replaced with the much larger $Z$ mass, and the momentum transfer is negligible.
Nevertheless, the differences do not translate into a very different energy deposition in the detector. To show this explicitly we have simulated both dark matter and neutrino interactions. The probability to get a hadronic shower with a given energy, for a fixed value of the energy of the incident particle (either a neutrino or a dark matter fermion) is shown in Fig.~\ref{fig:jetPDF}. 
As expected, the neutrino recoil energy is somewhat harder. However, this is a subdominant effect, while the largest differences between signal and background will be those associated to production. 

In order to consider the optimal location for a detector and estimate the sensitivity to light dark matter, we should take into account production and detection together and compute the number of signal and background events that a detector would observe at an off-axis angle $\theta$. Heavier mediators will generally broaden the angular distribution for the dark matter particles exiting the target, therefore increasing the signal rates for off-axis locations. The angular distribution will also be different depending on whether the particle produced in the $Z'$ decay is a fermion or a scalar. 

\begin{figure}[t]
 \begin{center}
  \includegraphics[width=.7\textwidth]{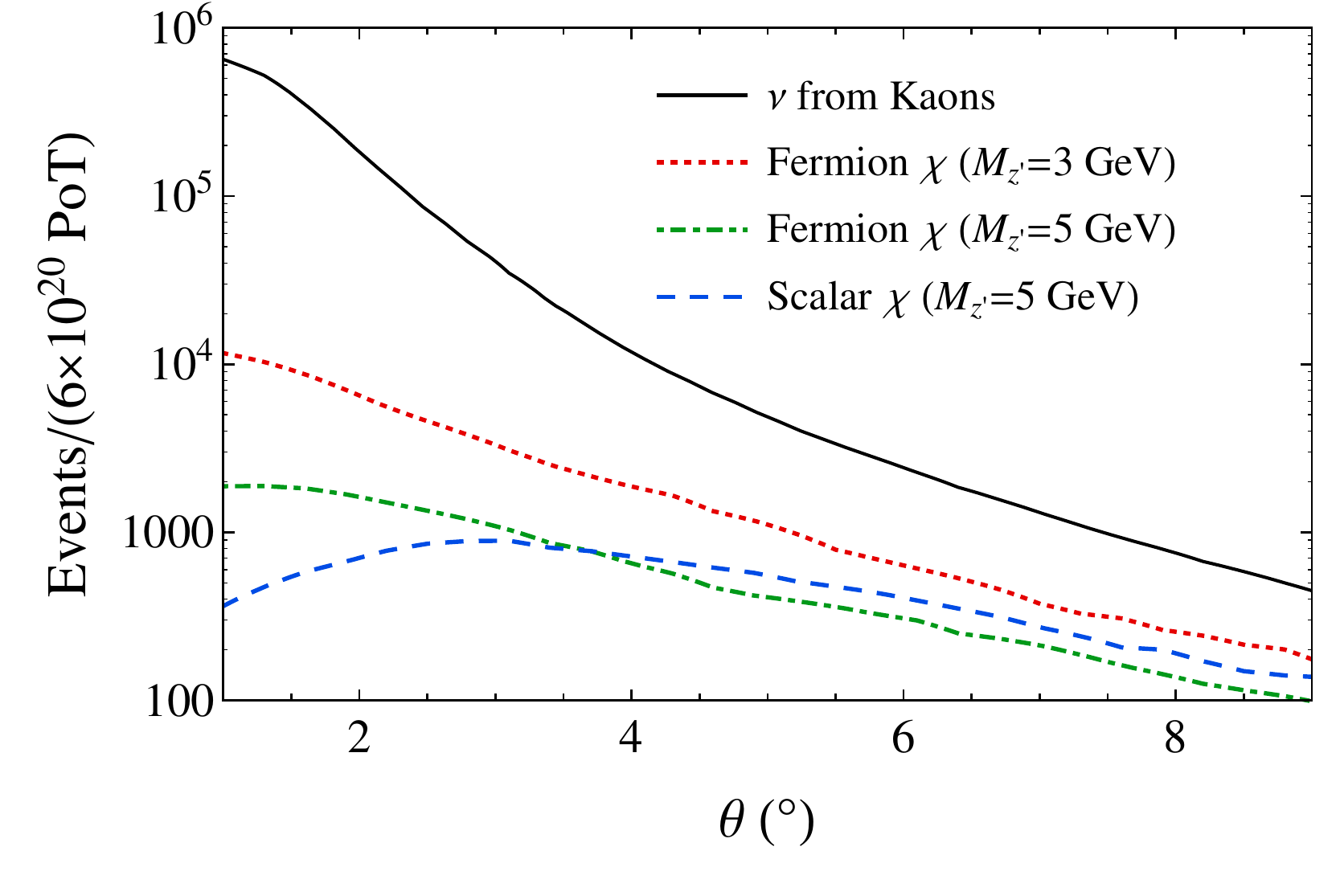}  
   \caption{Number of background (solid line) and signal (colored lines) events expected as a function of the off-axis angle, for a MiniBooNE-like detector located 745 meters away from the target  for 6$\times 10^{20}$ protons on target (PoT). 
   The dotted ($M_{Z'} = 3$ GeV)  and dot-dashed ($M_{Z'} = 5$ GeV) lines correspond to fermion DM, while the dashed line refers to scalar DM.  
     Here we assume $g_z= 0.1$, $z_\chi=3$ and $m_\chi = 750$~MeV.  }
   \label{fig:events}
 \end{center}
\end{figure}

The behavior of the total number of events with the off-axis angle is shown in Fig.~\ref{fig:events}, for the background as well as for three potential dark matter signals. The distance to the detector is fixed to $L=745$~m in this figure, and the angular acceptance of the detector is taken into consideration. As expected, the background falls much more rapidly  than the signals with the off-axis angle, which motivates to put the detector a few degrees off-axis. The effect of the heavier mediator mass can be seen from the comparison between the dotted and dot-dashed lines, while the effect of the spin of the dark matter particle is clearly seen from the comparison between the dashed and dot-dashed lines. As can be seen from the figure, the effect coming from the spin of the produced particle is the dominant. As expected, in the scalar scenario, more off-axis locations are clearly preferred, while if the dark matter particle is a fermion the preference is not as strong. The effect of the $Z'$ mass is subdominant.

\section{ Optimal detector location and expected sensitivity}
\label{sec:location}

\begin{figure}[t]
 \begin{center}
	  \includegraphics[width=.45\textwidth]{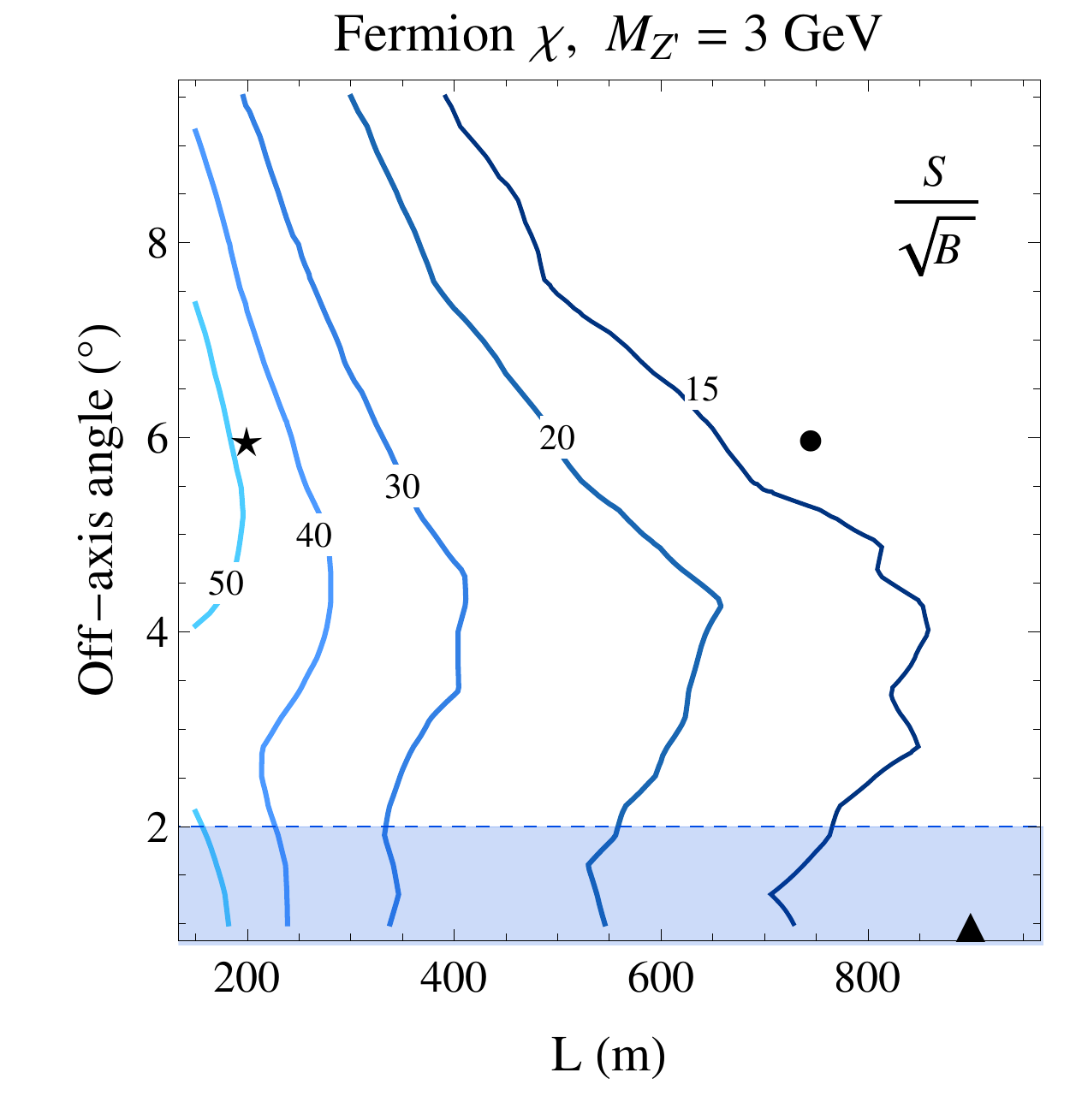} 
	  \includegraphics[width=.45\textwidth]{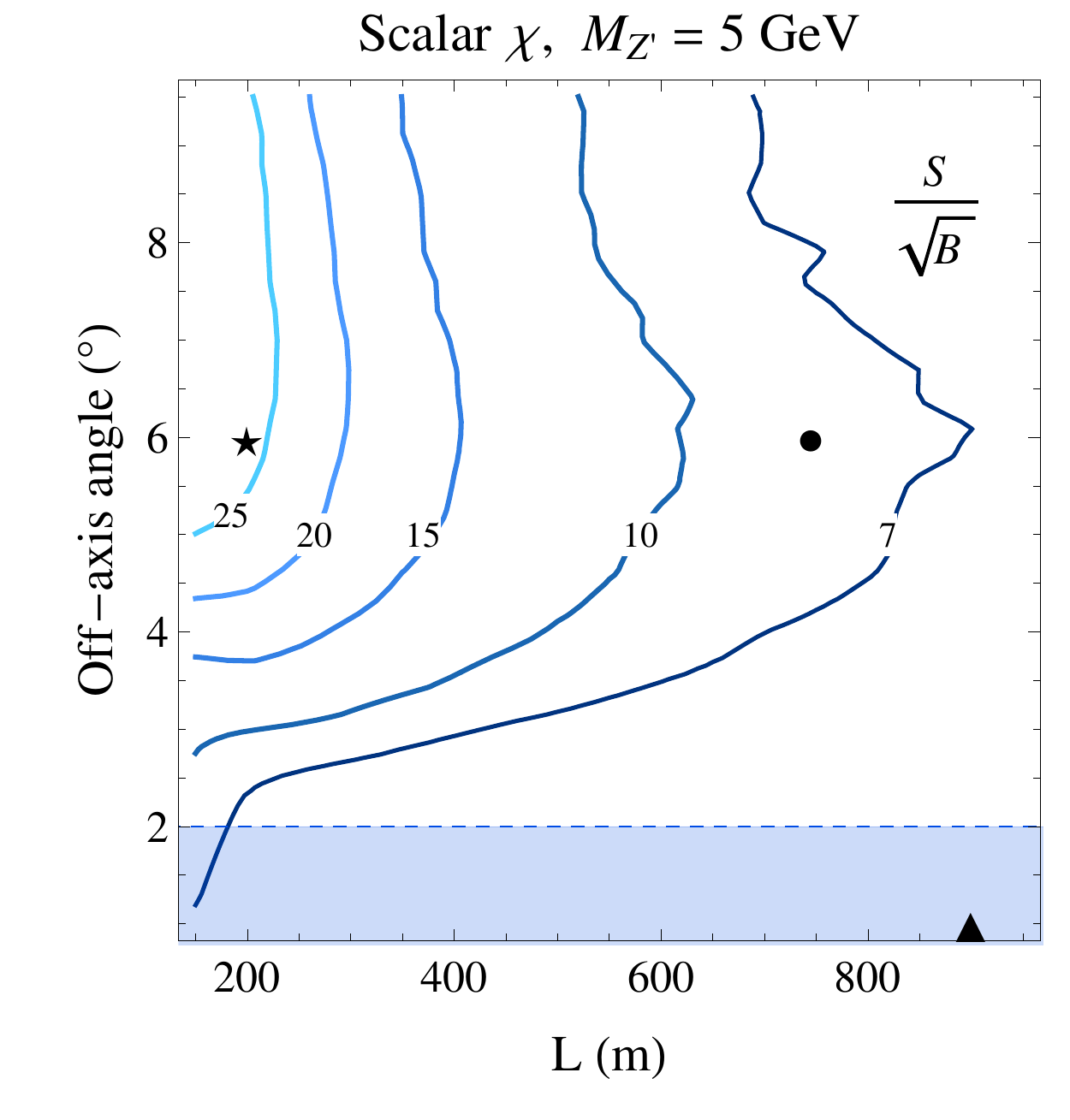} 
	\caption{ Contours of $S/\sqrt{B}$, where $S$ ($B$) is the number of DM signal (neutrino background) events as a function of the off-axis angle in degrees and the distance to the detector in meters,  for $g_{z}= 0.1$, $z_\chi=3$ and $m_\chi = 750$~MeV. 	
	 Left panel: signal significance  for a DM fermion  and $M_{Z'}=3$~GeV. Right panel: same for a DM scalar and $M_{Z'}= 5$~GeV.
 The triangle and the circle indicate the approximate locations of the NOvA near detector and MiniBooNE detector 
 from the NuMI target. The star marks the \emph{optimal} location for a dark matter search. 
	}
	\label{fig:optimal}
 \end{center}
\end{figure}

From the results shown in Sec.~\ref{sec:sig-bg} it is evident that, in order to achieve enough suppression of the neutrino background, an off-axis location for the detector is preferred. In this section, we make this statement more precise and determine the ideal location for a future LBNF detector to conduct a search for new light degrees of freedom coupled to the SM via a new vectorial force. For this purpose, we have computed the ratio between the total number of signal events ($S$) and the expected statistical uncertainty of the background event sample ($\sqrt{B}$), as a function of the off-axis angle and the distance to the detector. Our main result is summarized in Fig.~\ref{fig:optimal}, 
where the different lines correspond to iso-contours for particular values of $S/\sqrt{B}$, as indicated in the labels. The left panel shows the regions obtained for a $Z'$ with a mass of 3~GeV coupled to fermionic dark matter, while the right panel shows the results for a $Z'$ of 5~GeV coupled to a scalar particle. In both cases, the charge has been fixed to $z_\chi =3$, and the coupling is set to $g_z = 0.1$. A hypothetical ideal detector of approximately the MiniBooNE detector size has been assumed.

As expected from the results shown in Sec.~\ref{sec:sig-bg} (see also \cite{Dobrescu:2014ita}), the dependence with respect to the off-axis angle is different for the fermion and scalar cases. As can be seen from the plot, the ideal position of the detector in the scalar case with a heavier mediator shows a stronger preference for off-axis locations, while in the case of fermions it is less pronounced. It should also be noted that, since in the right panel the mediator chosen is heavier, the signal event rates will be consequently suppressed. Thus, the values shown in the contours for the $S/\sqrt{B}$ are lower in this case. 

In order to improve the sensitivity to light dark matter we must go further off-axis and study detectors that are not traditionally considered to be on the NuMI beamline. Our choice for an optimal detector is determined by the attempt of optimizing simultaneously the reach for both scalars and fermions. We therefore identify the ideal position (marked by a star) to be at roughly $ 6^\circ$ off-axis and at a distance of 200 m from the target, being the  minimal distance physically allowed by the presence of the decay pipe and focusing horn. Interestingly, the MiniBooNE detector (marked by a circle), which is on-axis with respect to the Booster beamline, is very close to the optimal off-axis angle identified in our study, although at a longer distance from the NuMI target ($L \sim 745$~m).

For reference, the approximate location of the NOvA near detector is indicated by a triangle in Fig.~\ref{fig:optimal}. As explained in the previous section, we only consider  neutrinos emitted from kaon decays as source of background. It should thus be kept in mind that, for angles close to the neutrino beam direction (\ie, for angles below 2$^\circ$ approximately) our computation may be underestimating the total number of background events. This is indicated in Fig.~\ref{fig:optimal} by the horizontal purple band. Just as an example, we checked that at the NOvA near detector about $10^6$ deep-inelastic scattering neutral-current events are expected when all neutrinos (coming both from $\pi$ and $K$ decays) are considered in the computation. This is an order of  magnitude above the result obtained when only neutrinos coming from $K$ decays are considered. From a similar argument it follows that the MINOS near detector would be even less sensitive to a possible light DM signal, being on-axis with respect to the neutrino beam.
 
As explained in Sec.~\ref{sec:model}, the model under consideration in this work contains a very small number of free parameters, namely: the coupling $g_z$, the charge of the dark matter under the $U(1)_B$ group, $z_\chi$, and the mass of the mediator between the SM and the hidden sectors, $M_{Z'}$. In this section, we will keep the value of $z_\chi$ fixed to $z_\chi = 3$, and determine the expected sensitivity to the coupling $g_z$, as a function of the mediator mass. Our results will be shown for the optimal detector location identified in Sec.~\ref{sec:location}, assuming an ideal detector of approximately MiniBooNE size, with perfect detection efficiency for neutral-current events.  For comparison, we will also show the expected results for the MiniBooNE detector location (always considering the NuMI target as the production point for the dark matter beam). It should be kept in mind that, since no special run would be needed to perform this search, an analysis could be done in principle using their past data\footnote{In fact, the MiniBooNE and MINOS collaborations have already published a joint measurement of the NuMI flux at the MiniBooNE detector~\cite{Adamson:2008qj}. } (including precise input about detector size and performance). 

In order to determine the sensitivity to the new coupling, a binned $\chi^2$ analysis is performed. The event rates are binned according to the energy deposited in the detector by the hadronic shower, using 1~GeV bins. In order to further reduce the background event rates, a minimum threshold of 3~GeV is imposed. A poissonian $\chi^2$ is then built as:
\begin{equation}
\chi^2 = \sum_i 2 \left\{ N_{bg,i} \ln \left(\frac{N_{bg,i} }{ N_{tot, i}(M_{Z'},g_z)} \right) + 
N_{tot, i} (M_{Z'}, g_z) - N_{bg,i}\right \} ,
\label{eq:chi2}
\end{equation}
where $N_{bg,i}$ stands for the background events in the $i$-th bin, and $N_{tot,i}$ stands for the total number of events expected in the same bin including the background plus a possible contribution from the signal (which depends on $M_{Z'}$ and $g_z$).

\begin{figure}[t!]
 \begin{center}
  \includegraphics[width=.55\textwidth]{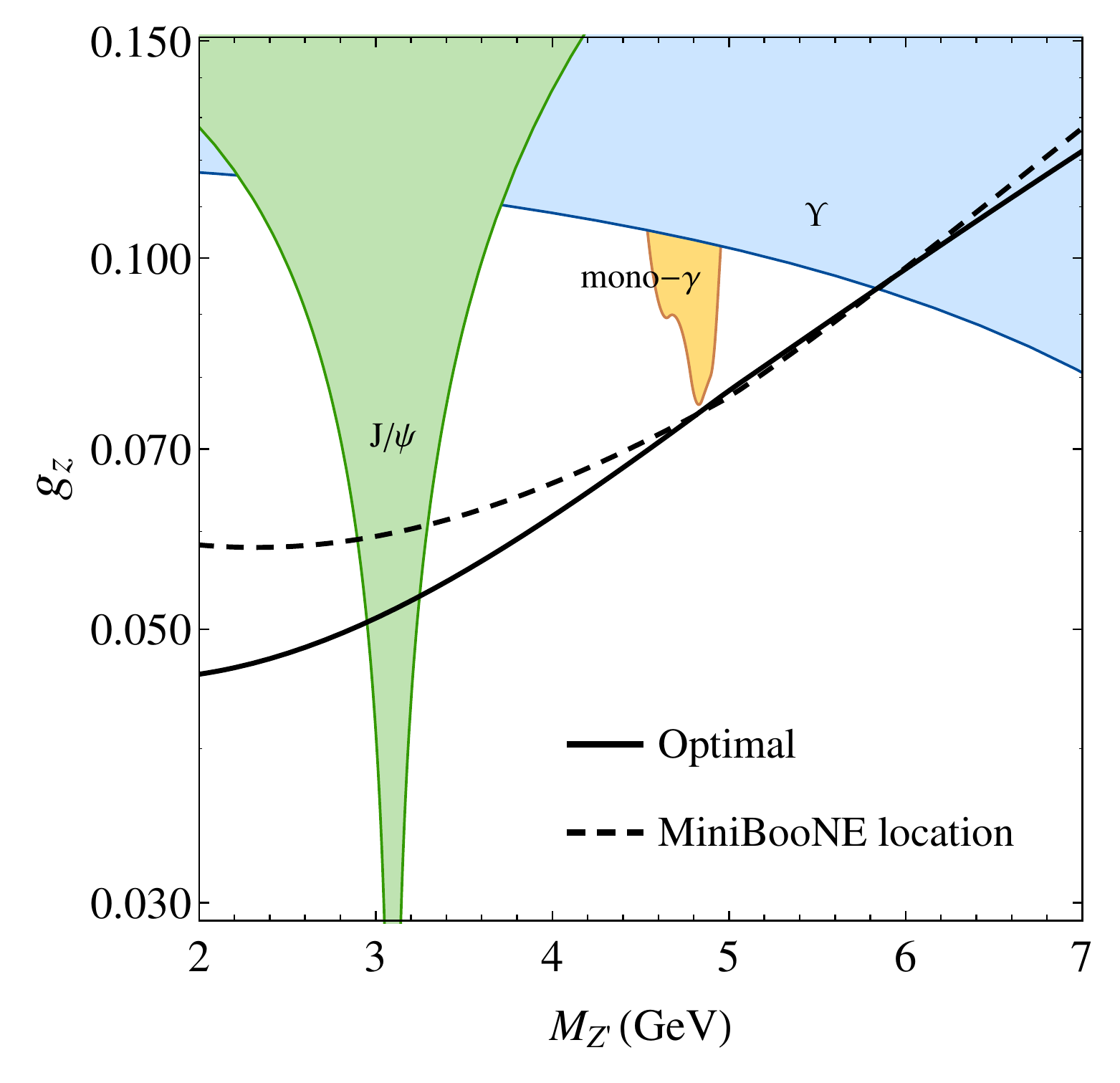} 
 \end{center}
 \caption{ Expected sensitivity (at 90\% C.L., 2 d.o.f.) to a DM fermion that interacts with quarks via a flavor-universal $Z'$ boson of mass $M_{Z'}$ and coupling $g_z$, for a
 DM $U(1)_B$ charge $z_\chi = 3$. The solid black line shows the sensitivity for a MiniBooNE-like detector at the optimal location from the NuMI target (see Fig.~\ref{fig:optimal}), while the dashed black line shows the sensitivity for a detector placed at the MiniBooNE/MicroBooNE site. The shaded areas are ruled out (see Sec.~\ref{sec:model}). 
 \label{fig:chi2}}
\end{figure}

The expected sensitivity contours are shown in Fig.~\ref{fig:chi2} for two possible detector locations: the optimal one (solid black lines) and the MiniBooNE location (dashed black lines). In both cases, a total exposure of 3.6$\times 10^{21}$ PoT has been considered. This corresponds to the nominal running time for the NOvA experiment of 6 natural years~\cite{Patterson:2012zs}. The contours are shown at the 90\% C.L. for 2 degrees of freedom (d.o.f.), and have been obtained assuming fermionic dark matter. For comparison, the strongest previous experimental bounds are also shown by the colored regions: monophoton searches at BaBar (yellow); and $J/\psi$ (green) and $ \Upsilon $ (blue) invisible decay searches, as discussed in Sec.~\ref{sec:model}.

For simplicity, no systematic errors have been considered when obtaining the $\chi^2$ contour. These will depend on the detector performance, cross section uncertainties, flux uncertainties at the detector location, etc.. Nevertheless, due to the strong dependence of the signal event rates with the coupling ($S \sim g_z^6$, see Sec.~\ref{sec:sig-bg}), we expect the final $\chi^2$ contour to remain largely unaffected by background normalization uncertainties. A larger effect could come from the detector performance parameters (detection efficiencies, for instance), since the sensitivity of the experiment in this scenario would be largely limited by statistics. A more careful study by the experimental collaborations is therefore needed to determine the final sensitivity for the search proposed here.

\section{ Conclusions}
\label{sec:concl}

The NuMI and LBNF neutrino beams rely on high-intensity proton fixed target facilities, with proton energies around 100~GeV, which can also be exploited to search for new light degrees of freedom. In particular, they could be essential to search for dark matter particles with masses below a few GeV, inaccessible at conventional direct detection experiments. The reason is that, if such dark matter particles exist and interact with nucleons, then a dark matter beam could be directly produced during proton collisions at the NuMI or LBNF targets. The subsequent dark matter detection would require a detector sensitive to neutral-current events, placed within a few hundred meters from the target. For a signal of this kind, though, neutrinos constitute the most relevant background. In this work we have investigated how it can be reduced. 

We have concentrated here on a scenario where both quarks and dark matter particles interact with a $Z'$ boson of mass in the  $1-10$ GeV range. The existing constraints on a $Z'$ boson of this type are loose, allowing its gauge coupling to be as large as 0.1. The $Z'$ can then be produced in large numbers at the LBNF, where its prompt decays into two dark matter particles would generate a wide beam. We have studied the dependence of the statistical significance of the signal with the off-axis angle and distance between the detector and the target. We have found that the ideal placement of a detector is at an off-axis angle of about $6^\circ$, and that a detector of the size of the MiniBooNE detector would be sensitive to a $Z'$ gauge coupling as low as 0.05. Our study motivates a proton beam at 120 GeV (or higher) in order to increase the sensitivity for models with a multi-GeV $Z'$ boson, resonantly produced at the target. It should be stressed that the strategy proposed in this work to search for dark matter can run symbiotically with the neutrino program, and a dedicated run would not be needed.  

We have also discussed the detection of a dark matter beam that may be produced in the NuMI beam line using existing detectors. The NO$\nu$A near detector would suffer from a large neutrino background due to the small off-axis angle. A similar argument would apply to the MINOS near detector. On the other hand, the detectors placed along the Booster beamline, such as MiniBooNE, MicroBooNE and possibly ICARUS, coincidentally subtend an ideal angle with respect to the NuMI beamline in order to conduct these searches. The lessons from this are twofold. First, the existing data set from MiniBooNE may be used to probe new regions of the parameter space in dark matter models. Second, this reveals strong synergies between the long- and short-baseline neutrino programs regarding new physics searches, which should be exploited and maximized in the future.

\bigskip

\acknowledgments

We are grateful  to Zarko Pavlovic for providing useful input regarding the kaon distributions and NuMI fluxes, as well as for numerous discussions. We thank Olivier Mattelaer for his help with \texttt{MadGraph} simulations, Zelimir Djuric for providing the neutrino fluxes at the NOvA near detectors, and Andr\'e de Gouvea, Lisa Goodenough, Raoul Rontsch and Sam Zeller  for useful discussions.  
We would also like to thank Roberto Vidal for writing a python parser~\cite{pylhef} for LHE files.

PC acknowledges financial support by the European Union through the ITN INVISIBLES (Marie Curie Actions, PITN-GA-2011-289442-INVISIBLES), and would like to thank the Mainz Institute for Theoretical Physics for hospitality and partial support during completion of this work. Fermilab is operated by Fermi Research Alliance, LLC under Contract No. \protect{DE-AC02-07CH11359} with the United States Department of Energy. 

\bigskip\bigskip\bigskip


\appendix 
\addcontentsline{toc}{section}{Appendix: Computation of the neutrino flux from kaon decays} 
\addtocontents{toc}{\protect\setcounter{tocdepth}{0}} 
\section*{Appendix: Computation of the neutrino flux from kaon decays}\label{sec:app:nuflux}
\renewcommand{\theequation}{A.\arabic{equation}}

As mentioned in Sec.~\ref{subsec:production}, the dark matter flux entering a detector placed at an off-axis location with respect to the beam direction is relatively easy to compute, since the $Z'$ is emitted very forward and to a good approximation its direction is the beam axis. However, the case of neutrinos being produced from kaon decays is very different, since kaons are typically produced at the target together with other hadrons which balance their $p_T$. Thus, kaons generally subtend a non-zero angle with respect to the beam direction, which has to be accounted for when computing the neutrino flux entering the detector.

The kaon energy and momenta distributions have been obtained from publicly available data in Refs.~\cite{Adamson:2008qj,Pavlovic:2008zz,zarko}. They were derived from a Monte Carlo simulation of the NuMI target, when exposed to a 120~GeV proton beam. Given this distribution of kaons, what is the neutrino distribution?
Since kaons decay relatively promptly, it is a good approximation to consider that all kaons decay at the beginning of the decay pipe. We will denote by $\theta_K, \phi_K$ the polar coordinates of the kaon in the lab frame, where $\theta_K$ is the polar coordinate with respect to the z-axis (which we choose to be the beam direction), and $\phi_K$ corresponds to the angle for a rotation in the x-y plane around the z-axis. 

It is important to recall that the angular distribution of neutrinos produced from a kaon decay with energy $E_K$ and momentum $\beta_K$ only depends on the kaon energy and on the neutrino angle with respect to the kaon momentum, $\theta_\nu$:
\[ 
\frac{dP}{d\Omega_\nu} = \frac{1}{4\pi}\frac{m_K^2}{E_K^2(1 - \beta_K \cos\theta_\nu)^2}\, .
\]
Moreover, the energy of a neutrino coming from a kaon with energy $E_K$ is: 
\begin{equation}
E_\nu (E_K, \theta_\nu) = \frac{m_K^2 - m_\mu^2}{2E_K(1 - \beta_K \cos\theta_\nu)}\, .
\label{eq:EnuEk}
\end{equation}
Therefore, for a fixed angle between the neutrino and the kaon rest frame, the energy of the neutrino is automatically determined by the kaon momentum. 

The computation of the total number of neutrinos produced from kaon decays that will reach the detector can be written as:
\begin{equation}
N_\nu = \int_{E_K, \Omega_K, \Omega_\nu} N_K (E_K,\theta_K) \frac{dP}{d\Omega_\nu}(E_K,\theta_\nu) d\Omega_K d\Omega_\nu dE_K \label{eq:Nnu}
\end{equation}
Here, $N_K (E_K,\theta_K)$ corresponds to the number of kaons with energy $E_K$ and angle $\theta_K$ which are produced in the target and decay producing a neutrino, and are extracted from a binned histogram given by the Monte Carlo simulation in Refs.~\cite{Adamson:2008qj,Pavlovic:2008zz,zarko}. Thus, the two integrals in $\theta_K$ and $E_K$ can be replaced by a discrete sum. Moreover, the kaon-neutrino system has a symmetry around the lab frame z-axis, so integration over $\phi_K$ only affects the overall normalization by a factor $\phi^{det}_K/\pi$, where $\phi^{det}_K$ is the aperture of the detector in the $\phi_K$ coordinate. 

In order to obtain the number of neutrinos reaching the detector, the integration limits have to be chosen according to the aperture of the detector. In particular, once the detector shape is considered, the limits on $\phi_\nu$ will depend on the value of $\theta_\nu$. Both neutrino coordinates in the lab system will also depend on the value of $\theta_K$. The integration in $\phi_\nu$ can be performed directly, and we are left with a function which depends on $\theta_\nu$ and the kaon variables. Therefore, Eq.~(\ref{eq:Nnu}) can be rewritten as:
\begin{equation}
N_\nu = \frac{\phi^{det}_K}{\pi} \sum_{E_K, \theta_K} \int^{\theta_{\nu}^{max}(\theta_K)}_{\theta_{\nu}^{min}(\theta_K)} \int^{\phi_{\nu}^{max}(\theta_K, \theta_\nu)}_{\phi_{\nu}^{min}(\theta_K, \theta_\nu) } N_K (E_K, \theta_K) \frac{dP}{d\Omega_\nu}(E_K,\theta_\nu) \sin\theta_\nu d\theta_\nu d\phi_\nu \, .
\end{equation}

In order to determine the integration limits, we have to take into account that the angular aperture of the detector is defined in the variable $\alpha$, which can be expressed as a function of the kaon and neutrino angular coordinates as:
\begin{equation}
\cos\alpha = - \sin\theta_\nu \cos\phi_\nu \sin\theta_k + \cos\theta_\nu \cos\theta_k \, .
\label{eq:alpha}
\end{equation}
This defines $\alpha$ as the angle between the neutrino produced in the decay and the beam (or z-) axis in the lab frame. In principle, the simplest solution would be to add a Heavyside function inside the integral, in such a way that the integrals in $\theta_\nu$ and $\phi_{\nu}$ are only performed for those values of $\theta_\nu$ and $\phi_\nu$ which satisfy the angular cut on $\alpha$. We found this to be computationally rather expensive, though.

Instead, we opted for the following approximation. For a very thin binning in the neutrino energy, the interval of allowed values of $\theta_\nu$ which give a neutrino inside the bin is very narrow, and much smaller than the aperture of the detector. Therefore, it can be easily checked whether the values of $\theta_\nu$ in this interval give a value of $\alpha$ inside the aperture of the detector. Within this approximation, the integral in $\theta_\nu$ can be taken as the value of the function in the middle of the integrating interval, times the size of the interval. Also, the integrand does not depend on $\phi_\nu$ anymore and can be integrated independently. As a result, we get:
\begin{equation}
N_\nu (E_{\nu,i}) \simeq \frac{\phi^{det}_K}{\pi} \sum_{E_K}\sum_{\theta_K} N_K (E_K, \theta_K) \frac{dP}{d\Omega_\nu}(E_K,\overline{\theta_\nu}) \sin\overline{\theta_\nu} \cdot \Delta \theta_\nu \cdot \Delta\phi_\nu(\overline{\theta_\nu}, \theta_K)  \, ,
\label{eq:finalapprox}
\end{equation}
where $N_\nu(E_{\nu,i})$ now corresponds to the number of neutrinos entering the detector with energies inside the $i$-th neutrino energy bin, and 
 \begin{eqnarray}
 \Delta \theta_\nu & = & \theta_\nu^{max} - \theta_\nu^{min} \nonumber \, , \\
 \overline{\theta_\nu} & = & \theta_\nu^{min} + \Delta \theta_\nu/2  \, ,  \\
 \Delta \phi_\nu (\overline{\theta_\nu}, \theta_K) & = & \phi_\nu^{max} (\overline{\theta_\nu}, \theta_K) - \phi_\nu^{min} (\overline{\theta_\nu}, \theta_K) \, . \nonumber 
 \end{eqnarray}
 
Before computing the contribution to the neutrino flux for a given energy bin by using Eq.~(\ref{eq:finalapprox}), though, the acceptance condition in $\alpha$ is required to be satisfied, \textit{i.e.,} it is required that the interval of $\theta_\nu$ corresponding to each neutrino energy bin gives an interval in $\alpha$ inside the angular acceptance of the detector. 

We find that, under these approximations, our computation of the kaon contribution to the total neutrino flux shows a good agreement with the fluxes from Ref.~\cite{Adamson:2008qj} for a MiniBooNE-like detector located $6^\circ$ off-axis and at a distance of $L=745$~m from the source.


\providecommand{\href}[2]{#2}\begingroup\raggedright\endgroup

\end{document}